\documentclass[12pt]{article}
\usepackage{psfig}
\begin{document}
\title{Charmonium Suppression with $c\bar{c}$ Dissociation 
by Strings\footnote{Supported by BMBF and GSI Darmstadt}}
\author{J.~Geiss, C.~Greiner, E.~L.~Bratkovskaya, W.~Cassing and U.~Mosel\\
Institut f\"ur Theoretische Physik, Universit\"at Giessen \\
D-35392 Giessen, Germany}
\date{ }
\maketitle

\begin{abstract}
We study the production of $c \bar{c}$ pairs in nuclear reactions at
SPS energies within 
the covariant transport approach HSD. The production of $c \bar{c}$ is treated
perturbatively employing experimental cross sections while the
interactions of $c\bar{c}$ pairs with baryons are included by
conventional cascade-type two-body collisions. Adopting 6~mb for the $c
\bar{c}$-baryon cross sections the data on $J/\Psi$ suppression in
p~+~A reactions are reproduced in line with calculations based on the
Glauber model. Additionally the dissociation of the $c \bar{c}$ pairs by strings
is included in a purely geometrical way. 
We find good agreement with experimental
data from the NA38 and NA50 collaboration with an estimate for the string 
radius of $R_s \approx 0.2-0.3 \,fm$. 
\end{abstract}

\vspace{0.4cm}
\noindent

\newpage
\section{Introduction}

The $J/\Psi$ suppression has been proposed as one of the signals
for the quark-gluon plasma (QGP) which is expected to be formed in
nucleus-nucleus collisions at sufficiently high energies \cite{matsui}.
This suggestion has stimulated a number of experiments,
which indeed have  observed a significant reduction of the scaled
$J/\Psi$ yield when going from proton-nucleus to nucleus-nucleus
reactions \cite{NA38}. Especially the NA50 experiment has reported an
abrupt decrease in  $J/\Psi$ production in Pb~+~Pb collisions
at 158 GeV per nucleon \cite{gonin,NA50,NA50a} in going from peripheral to central
collisions.

A lot of theoretical effort has been spent 
to understand the experimental results (see \cite{gerhuf} for
a recent review). 
Beside the suggestion of a possible formation of a QGP 
\cite{KhaQM96,blaizot,wong} various scenarios based 
on $J/\Psi$ absorption by hadrons have also been proposed 
\cite{gerschel,gavin,gavin1,Capella,cassing-jpsi,armesto}. 
A part of the suppression can  be explained
by $J/\psi $--absorption on the surrounding nucleons
\cite{gerschel}, additional absorption
might be attributed to `comovers' (`mesons')
being produced as secondaries \cite{gavin,gavin1}.
It has been shown in \cite{cassing-jpsi}
within the microscopic covariant 
transport approach HSD (Hadron String Dynamics) \cite{cassing} that the observed
suppression of the $J/\Psi$ yield in nuclear collisions is consistent with such 
an hadronic absorption scenario.
However, the 'comover' models are still a matter of debate: The employed
absorption cross section of $c\bar{c}$ pairs on comovers 
of 3 mb is treated as a free  parameter in order to explain the data.
On the other hand, this cross section might be overestimated 
considerably \cite{antico}.

In our present work we therefore consider an alternative mechanism 
for  $J/\Psi$ production in heavy ion collisions.
We will focus on the effect of $c\bar{c}$ dissociation in the prehadronic phase
and not on absorption by comovers.
This is motivated by the fact, that the very early collision phase is 
not described by hadrons but by highly excited strings, which in the HSD model 
are created by momentum transfer among target and projectile
nucleons . Their 
production and decay is treated within the FRITIOF model \cite{LUND} and
describes the first few fm/c of the collision. The $c \bar{c}$ state, most likely
as a color octet, is also produced at the earliest state of the
reaction in a hard collision among the nucleons via gluon-gluon fusion. 
Hence it is natural to ask 
what happens if this colored state moves into such a temporary 
environment of strings. 
As the string carries a lot of internal energy (to produce the later secondaries)
in a small and localized space-time volume
the quarkonia state might get completely dissociated by the
intense color electric field inside a single string. 
In this 
respect Loh et. al.   \cite{loh} have investigated the possible $J/\Psi$ 
dissociation in a color electric flux tube in a semiclassical model
based on the Friedberg-Lee color dielectric Lagrangian. They find that 
the $c \bar{c}$ state becomes dissociated rather immediately on a timescale
less than 1 fm/c. 
In a heavy ion collision, especially for the more
heavy systems, the effective region (or volume) of all the strings
being produced is expected to be large so that the strings might 
become rather closely packed.
If the $c\bar{c}$ states will get dissociated by the numerous
and individual strings this will lead to an additional suppression
of $J/\Psi$ in the early phase prior to hadronization. 
To explore this intuitive idea we include the effect of
$c\bar{c}$ dissociation by strings into the HSD model and study the overall
$J/\Psi$ production and dissociation dynamically.

The HSD model is briefly presented in section 2, while  in section 3 we describe 
the $c\bar{c}$ production process and show details of the string
evolution in  the HSD model. We then address the dissociation process
of the $c\bar{c}$ states on baryons and the chromoelectric flux tube of
the strings (section 4). We conclude with a summary of our investigations.

\section{The covariant transport approach}
In this work we perform our analysis along the line of the HSD approach
\cite{cassing} in the cascade modus 
which is based on a coupled set of covariant transport
equations for the phase-space distributions $f_{h} (x,p)$ of hadron $h$
\cite{cassing}, i.e.
\begin{samepage}
\begin{eqnarray}  \label{g24}
\lefteqn{\left( {\partial \over \partial t} + {\vec{p}_1 \over m} 
\vec{\nabla} \right)
f_1(x,p_1)  } \nonumber \\
&&= \sum_{2, 3, 4\ldots}\int d2 d3 d4 \ldots
 [G^{\dagger}G]_{12\to 34\ldots}
\delta^4(p_1^{\mu} +p_2^{\mu} -p_3^{\mu} -p_4^{\mu}\ldots )  \nonumber\\
&& \times \left\{ f_3(x,p_3)f_4(x,p_4)\bar{f}_1(x,p)
\bar{f}_2(x,p_2)\right.  \nonumber\\
&& -\left. f_1(x,p)f_2(x,p_2)\bar{f}_3(x,p_3)
\bar{f}_4(x,p_4) \right\} \ldots\ \ .
\end{eqnarray}
\end{samepage}
Here $ [G^{\dagger}G]_{12\to 34\ldots}
\delta^4(p_1^{\mu} +p_2^{\mu} -p_3^{\mu} -p_4^{\mu}\ldots )$ 
is the `transition rate' for the process
$1+2\to 3+4+\ldots$,
while the phase-space factors
\begin{equation}
\bar{f}_{h} (x,p)=1 \pm f_{{h}} (x,p)
\end{equation}
are responsible for fermion Pauli-blocking or Bose enhancement,
respectively, depending on the type of hadron in the final/initial
channel. The dots in eq.~(\ref{g24}) stand for further contributions in
the collision term with more than two hadrons in the final/initial
channels. The transport approach (\ref{g24}) is fully specified 
by the transition rates $G^\dagger G\,\delta^4 (\ldots )$ in the collision
term, that describes the scattering and hadron production and absorption
rates. This transport approach
was found to describe reasonably well hadronic as well as dilepton
data from SIS to SPS energies \cite{cassing,cassing1}.

\begin{figure}[h1]
\centerline{\psfig{figure=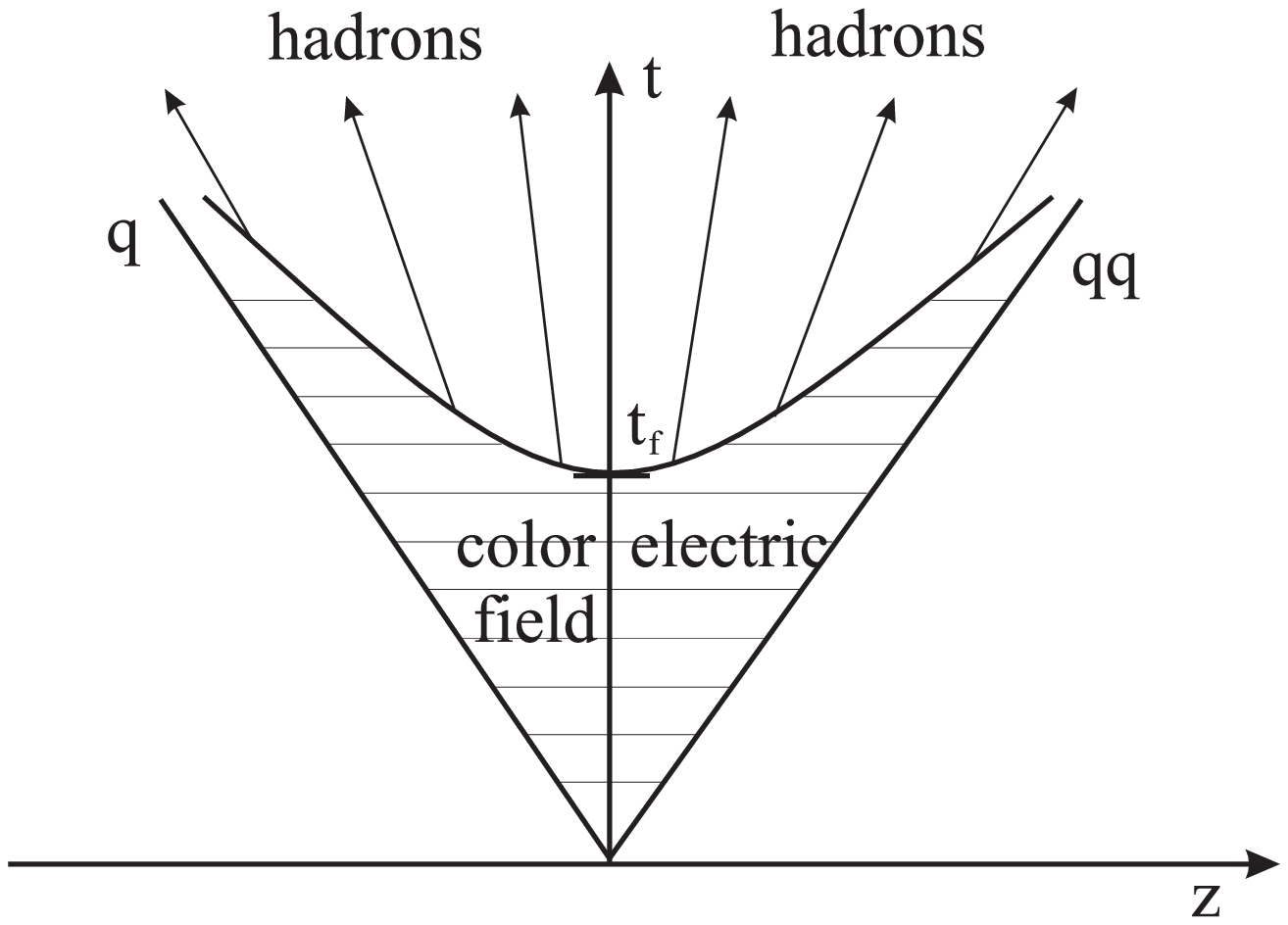,height=10cm}}
\caption{Dynamical evolution of a baryonic string. The fragmentation
starts after the formation time $t_f$.}
\label{string_dyn}
\end{figure}

In the present approach we propagate explicitly -- apart from the
baryons (cf.~\cite{cassing}) -- pions, kaons, $\eta$'s, $\eta^\prime$'s,
the $1^-$ vector mesons $\rho, \omega, \phi$ and $K^*$'s as well as the
axial vector meson $a_1$. 
The high energy hadron-hadron 
collisions are described by the FRITIOF model \cite{LUND} 
resulting in two excited 
strings. The dynamical evolution of the strings is now
included explicitly. A string is characterized by the leading quark and
diquark (or antiquark in the case of a mesonic string) and by the energy 
stored in between. 
The fragmentation of the strings into hadrons starts after the 
formation time, which is set to  $\tau_F=0.8$ fm/c 
(see Fig. \ref{string_dyn}). $\tau_F$ controls the baryon and meson 
rapidity  distribution $dN/dy$ in comparison to experimental data.
The length of the strings is given by the (center of mass) 
collision time $t_0$  of the 
hadrons, the formation time and the velocity of the
leading quarks/diquarks $\beta_i=|\vec{p}_i|/E_i$, $i=1,2$
\begin{eqnarray}  \label{stringlength}
l(t)=\left\{ 
\begin{array}{lcl}
|\beta_2-\beta_1| \cdot (t-t_0) &, & (t-t_0) \le \tau_F  \\
\left(|\beta_2-\beta_1| - 2 \cdot \sqrt{1-{\tau_F^2 \over (t-t_0)^2}} \right) 
\cdot (t-t_0) &, & \tau_F < (t-t_0) \le t_{max} \\
0 &, & (t-t_0) > t_{max},  
\end{array}
\right. 
\end{eqnarray}
where $t_{max}=t_0+2 \tau_F / (\sqrt{ 4 - |\beta_2-\beta_1|^2})$ is the time
when the string is completely hadronized.
The radius of the flux tube is an unknown 
parameter, which will be of particular significance for the $c\bar{c}$ dissociation 
by strings, which we will address in section 4. 
The cross section of the high energy secondary interactions of the leading 
quarks/diquarks are treated within a simple additive quark model 
$\sigma(q-B)=1/3 \,\sigma(B-B)$ and $\sigma(qq-B)=2/3 \, \sigma(B-B)$. 

\begin{figure}[h1]
\hspace{-1.5cm}
\psfig{figure=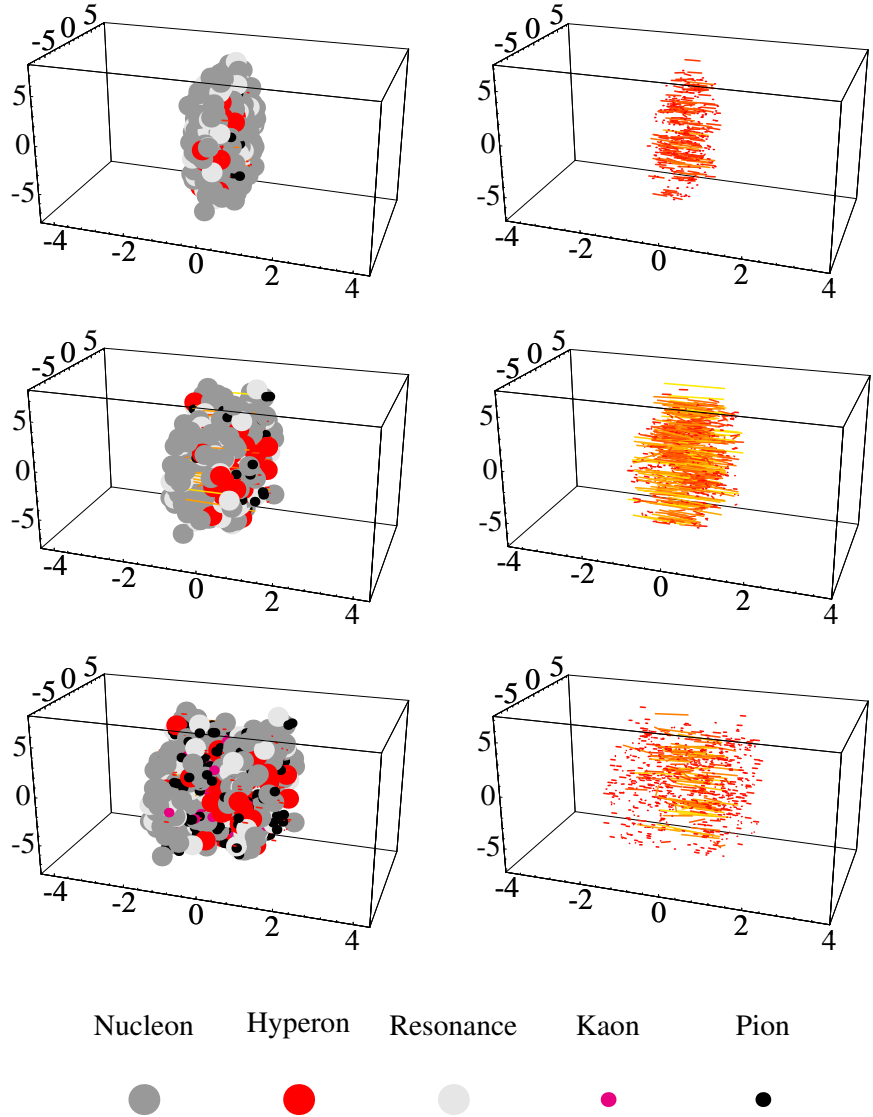,height=18cm}
\caption{Graphical representation of the hadrons (left) 
and the strings (right)
during the high density phase in a central 
Pb~+~Pb collision at 160 AGeV in the center of mass system.
Three time steps are shown at $t_{cm}=(4.4, 5.0, 5.6)$ fm/c 
(see next figure); 
the axis labels are given in fm. }
\label{string-pic}
\end{figure}

\begin{figure}[h1]
\centerline{{\psfig{figure=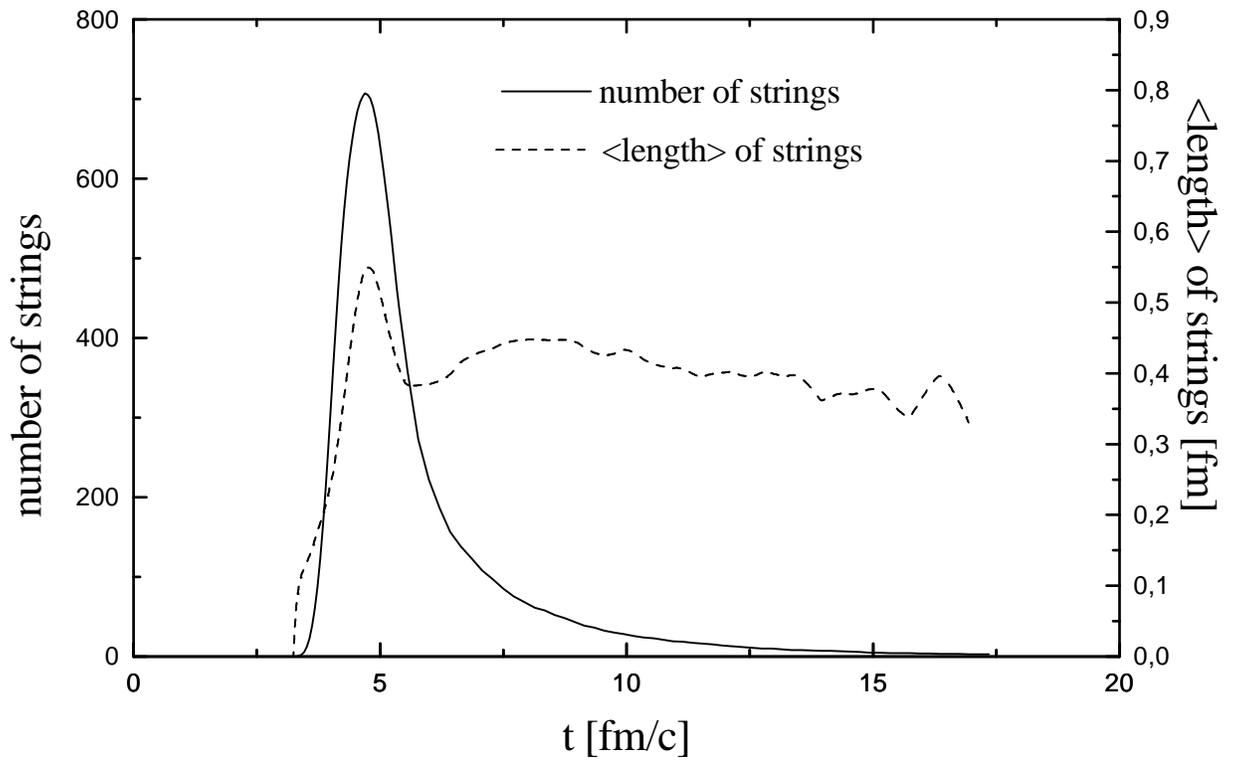,height=12cm}}}
\caption{Number of strings and  average string length 
in a central Pb-Pb collision at 160GeV as a function of time in the 
nucleus-nucleus cms.}
\label{string_number}
\end{figure}

In our simulation the production and decay of strings dominates the very early 
collision phase before the subsequent hadronic state of matter is formed. 
In Fig. \ref{string-pic} a characteristic representation of the hadrons  
and the strings during the high density phase in a central 
Pb~+~Pb collision at 160 AGeV is shown. 
As pointed out already in the introduction several
hundred strings are formed during a central Pb-Pb collision at SPS energies.
It turns out that most of them are in fact rather short due to 
secondary interactions of the 
leading quarks/diquarks. In Fig. \ref{string_number}
the number of strings and  the averaged string length 
in a central Pb-Pb collision at 160GeV is shown as a function of 
the center of mass time.

After the hadronization many low energy rescattering processes
are included. The  meson-baryon and baryon-baryon collisions are treated 
similar to the BUU model \cite{Wolf90}.
As meson-meson channels we include the reactions $\pi \pi \rightarrow \rho,
\pi \pi \rightarrow K \bar{K}, \pi \rho \rightarrow \phi, \pi \rho
\rightarrow a_1$ as well as the time reversed reactions using Breit-Wigner
cross sections with parameters from the literature \cite{PDB} and
exploiting detailed balance. For the present analysis the low
energy rescattering processes do not play an essential role. 
Within the philosophy outlined in the introduction the production
and absorption of the $c\bar{c}$ states happens in the first few fm/c
of the collision before secondary particles are produced.

\section{Charmonium production}

Since the probability of producing initially a $c \bar{c}$ pair is
very small, a perturbative approach is used for technical reasons
as described in \cite{cassing-jpsi}.
Whenever  two nucleons collide a $c\bar{c}$ pair is produced with a
probability factor $W$, which is given by the ratio of the $J/\Psi$
 to $NN$ cross section at a center-of-mass energy
$\sqrt{s}$ of the baryon-baryon collision,
\begin{equation}
W= \frac{\sigma_{BB \rightarrow J/\Psi + X} (\sqrt{s})}
{\sigma_{BB \rightarrow BB+X} (\sqrt{s})}.
\label{wratio}
\end{equation}
The parameterization used for the $J/\Psi$ cross section, the rapidity
and $p_t$ distribution of the $c\bar{c}$ states is taken as in
\cite{cassing-jpsi}, which fits reasonable well the experimental data.

It is known from experiment that the Drell-Yan cross section as a
hard process scales with $(A_P \times A_T)$ \cite{NA38,NA50,Carlos}.
This is in contrast to most of the soft hadronic observables, like
e.g. the pion multiplicity, which scale like $A_P + A_T$. This indicates
a strong difference in the production of hard and soft processes in
heavy ion collisions. 
Since the production scheme for $c\bar{c}$ is similar to the production 
of Drell-Yan pairs the total initial $c\bar{c}$ production 
cross section also scales with $A_P \times A_T$. It
is important to stress that only with this stringent assumption 
the ratio of the $J/\Psi$ to the Drell-Yan cross section is a direct measure
for the  $J/\Psi$ suppression. We want to note that the 
$A_P \times A_T$ behavior of the initial
$c\bar{c}$ production cross section is still a matter of debate. 
As pointed out by Frankel and Frati \cite{Frankel} and also by Tai An et al. 
\cite{Tai} the strong $J/\Psi$  suppression in central Pb+Pb collisions might be 
explained by a nonlinear  scaling of the 
initial $c\bar{c}$ production cross section due to initial state 
interactions. Such initial state interactions should also diminish the
Drell-Yan cross section, which, on the other side, is not seen
experimentally (for a comprehensive discussion on this issue we
refer to \cite{kharz}).
In any case, in our approach we stay with the same scaling behavior 
and investigate 
the subsequent absorption of $c\bar{c}$ pairs on baryons and strings. 
For that reason we emphasize 
that the $A_P \times A_T$ scaling of the initial $c\bar{c}$ production
is an input of our calculation and not a result. To implement this scaling we
separate the production of the  
hard and soft processes: The space-time production vertices of the 
$c\bar{c}$ pairs are calculated before every run neglecting the soft
processes. After that we follow the motion of the Charmonium pairs  
within the full hadronic background by propagating it as a free particle. 
Again, only with
this concept the $A_P \times A_T$ dependence of the $c\bar{c}$ production
can be reproduced in a hadronic transport simulation.

\section{$c\bar{c}$ dissociation}
The $c\bar{c}$ state as a rather heavy hadronic particle and a result 
of a hard process is formed immediately 
in comparison to the soft particle production
from string fragmentation. Thus first of all the $c\bar{c}$ states 
move not in a hadronic environment but in an environment of color electric
strings of 'wounded' nucleons (see Fig. \ref{string-pic}). 
As motivated in the introduction and in ref.\cite{loh} we now 
assume that a $c\bar{c}$ state immediately
dissociates whenever it moves into the region of the color electric 
field of a string. In this sense strings are completely 
black for $c\bar{c}$ states. It was found in \cite{loh} that the additional force
acting on the charm quarks is given by $2 \times \sigma \approx 2 \,GeV/fm$,
where $\sigma$ denotes the phenomenological string constant of a 
chromoelectric flux, which is sufficient to immediately break up a 
$c\bar{c}$ state. One can also argue that the field energy density contained in a 
string is given by $\sigma / (\pi R_S^2)$. For $R_S \approx 0.3 \,fm$ one
accordingly has a local high color electric energy density of 
$\approx 4\, GeV/fm^3$, which substantially screens the binding potential
of the Charmonium state \cite{matsui}. 

For practical reasons the dissociation by a string is modeled
when the center of mass of the $c\bar{c}$ state is located inside the string. 
The length of the strings is given by eq. (\ref{stringlength}), while 
the string radius $R_s$ is an unknown parameter. 
Absorption by  strings spanned between the parent particles of the 
$c\bar{c}$ pair are excluded, since this effect already 
is included in the production cross section.

Additionally, the $c\bar{c}$ pair may be destroyed by collisions with incoming
baryons. For the collisions with baryons we use the minimum
distance concept described in Ref. \cite{Wolf90}.  For the actual cross
sections employed we assume that the $c\bar{c}$ pair initially is 
produced in a color-octet
state and immediately picks up a soft gluon to form a color neutral
$c\bar{c}-g$ Fock state \cite{KhaQM96} (color dipole).  This extended
configuration in space is assumed to have a 6~mb absorption cross
section in collisions with baryons ($c \bar{c}+B\to\Lambda_c+\bar D$)
as in Refs.  \cite{gonin,gerschel,KhaQM96} during the lifetime $\tau$
of the $c\bar{c}-g$ state, for which we adopt $\tau =$ 0.3 fm/c as suggested
by Kharzeev \cite{KhaQM96}. One also has to specify the
absorption cross sections of the formed resonances $J/\Psi$ 
on baryons. For simplicity we use 3~mb following Refs.
\cite{KhSatz96,cassing-jpsi}.

\section{Results}
\begin{figure}[ht]
\centerline{{\psfig{figure=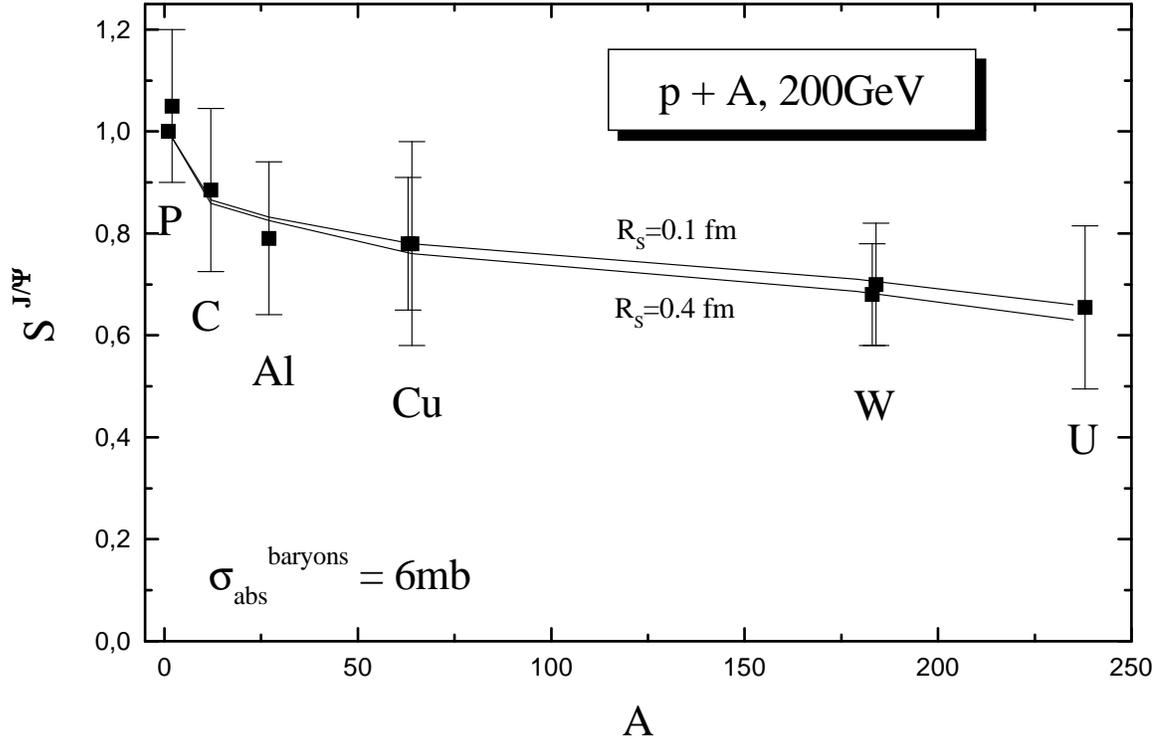,height=12cm}}}
\caption{The $J/\Psi$ survival probability $S^{J/\Psi}$ for 
minimum bias p~+~A reactions
at 200 GeV assuming a 6 mb cross section for the $c\bar{c}$ absorption
on baryons including $c\bar{c}$ dissociation on strings for two different
string radii, $0.1$ fm  and $0.4$ fm in comparison 
to the experimental data from \protect\cite{gonin}.}
\label{pa_abs}
\end{figure}

In proton-nucleus collision the $c\bar{c}$ dissociation by strings 
is practically negligible. The absorption is dominated by 
$c\bar{c}$-baryon collisions so that the proton nucleus data allow to 
fix this absorption cross section. In Fig. \ref{pa_abs} we show 
our results for the $J/\Psi$ survival probability  $S^{J/\Psi}$  
using 6~mb for the
absorption cross section of the $c \bar{c}$-pairs on nucleons in
comparison to the data \cite{gonin}.
The experimental `survival
probabilities'  $S_{exp}^{J/\Psi}$ in this figure as well as in the
following comparisons are defined by the ratio of experimental
$J/\Psi$ to Drell-Yan cross sections as
\begin{eqnarray}
S_{exp}^{J/\Psi} = \left.{\left(B_{\mu\mu}\sigma^{J/\Psi}_{AB}\over
\sigma^{DY}_{AB}|_{2.9-4.5 \ {\rm GeV}}\right)} \right/
{\left(B_{\mu\mu}\sigma^{J/\Psi}_{pd}\over \sigma^{DY}_{pd}\right)},
\label{sexp}\end{eqnarray}
where $A$ and $B$ denote the target and projectile mass while
$\sigma^{J/\Psi}_{AB}$ and $\sigma^{DY}_{AB}$ stand for the $J/\Psi$ and
Drell-Yan cross sections from $AB$ collisions, respectively, and
$B_{\mu\mu}$ is the branching ratio of $J/\Psi$ to dimuons.
The theoretical ratio is defined as
\begin{eqnarray}
S_{theor}^{J/\Psi} = {M_{J/\Psi}\over N_{J/\Psi}},
\label{stheor}\end{eqnarray}
where $N_{J/\Psi}$ is the multiplicity of initially produced $J/\Psi$'s
while $M_{J/\Psi}$ is the multiplicity of $J/\Psi$'s that survive
the hadronic final state interactions.

In Fig. \ref{pa_abs} results are shown for two different string radii
$R_S=0.1fm$ and $R_S=0.4fm$. The  difference of the
curves is rather small, which indicates the small effect of 
dissociation by strings.
For p~+~U and $R_S=0.4 fm$ only 2\% of the $J/\Psi$ are absorbed by strings. 
This is completely different for heavy ion collisions,
where the absorption on strings becomes a much more important effect. 

To compare our results for S~+~U and Pb~+~Pb to the NA38 and NA50 data,
the experimental trigger conditions must be included.
In these  experiments only events are recorded with a
$\mu^+\mu^-$ pair of invariant mass $M\ge 1.5$~GeV. This trigger
condition is obtained by 
\begin{eqnarray}
{d\sigma_{theor}^{\mu\mu} \over E_T} = 2\pi N_0 \int\limits_0^\infty
b db \ {dN\over dE_T}(b) \ \sum\limits_{i} W_i^{\mu\mu} (b),
\label{ET-theor}\end{eqnarray}
where $W_i^{\mu\mu} (b)$ are the weights for produced $\mu\mu$ pairs
within the experimental cuts. 
$N_0$ is a normalization factor to adjust to the experimental
number of events. We take the same weight factors $W_i^{\mu\mu} (b)$
as in \cite{cassing-jpsi} which reasonably reproduce the experimental
$E_t$ distribution as shown in \cite{cassing-jpsi}.

Qualitatively the results are not changed by this procedure. In Fig.
\ref{hi_sup} our results are shown for  S~+~U and Pb~+~Pb as a function of 
the transverse energy and for four different string radii. A strong dependence
on the string radius $R_S$ is observed and $R_S=0.2 fm$ gives the best fit 
to experimental data \cite{NA50,NA50a}. With this string radius 40\% of the absorbed 
$J/\Psi$'s are dissociated by strings in central collisions 
of Pb~+~Pb as shown in  Tab. \ref{tab1}.

\begin{table}
\centerline{\begin{tabular}{|c|c|c|c|}
\hline
$R_S [fm]$ & $A_{tot}$ & $A_B$ & $A_S$  \\
\hline\hline
0.1        & 0.65    &  0.53 & 0.12   \\ \hline
0.2        & 0.74    &  0.44 & 0.30   \\ \hline
0.3        & 0.82    &  0.36 & 0.46   \\ \hline
0.4        & 0.88    &  0.29 & 0.59   \\ \hline
\end{tabular}}
\caption{Total $J/\Psi$ absorption probability $A_{tot}=1-S_{theor}^{J/\Psi}$, 
absorption by baryons
$A_B$ and by strings $A_S$ for different string radii in a central 
Pb~+~Pb collision as obtained
in the HSD approach.}
\label{tab1}
\end{table}

\section{Summary}
In this work we have addressed the question of $c\bar{c}$ absorption
in relativistic heavy ion collisions in the microscopic 
transport approach HSD.
Of particular interest was the effect of the prehadronic phase, which 
in the HSD model is described by independent strings. 
In principle the comover absorption
scenario could be included in addition.
However, since most of the $c\bar{c}$ pairs are already absorbed by
strings and baryons in the early stage of the reaction, the effect
of comovers is expected to be much less than found in 
ref. \cite{cassing-jpsi}.

The dissociation mechanism by strings was treated in a 
simple geometrical picture. Whenever a 
$c\bar{c}$ pair moves into a string it dissociates and forms $D\bar{D}$ 
mesons. 
Strings are completely 'black' for $c\bar{c}$ states. 
This stringent condition is a first ansatz to investigate the effect
of the $c\bar{c}$ dissociation by strings. Better theoretical foundations
of the dissociation mechanism of charmonium ($J/\Psi$ and $\psi'$) and 
preresonant states ($c\bar{c}g$) are definitely needed
to improve the understanding of this geometrical picture. 

We found that the absorption by strings plays is an important effect in the 
first few fm/c of the collision phase before secondary particles
are produced. Adopting a string radius of $0.2 fm$ we got a qualitative
agreement with p+A data and the NA38 and NA50 data. This radius seems to
be rather small, but for two reasons it should be seen as a lower bound 
for $R_S$.
First of all, we have assumed that the strings are completely black for  
$c\bar{c}$ states, and secondly it 
dissociates whenever it moves into the region of the color electric 
field of a string. With the requirement
that the total $c\bar{c}$ state should be inside the string when the 
dissociation
process starts, one should add the $c\bar{c}$ radius to our value of $R_S$.
This would give a string radius of $\tilde{R}_S\approx 0.4-0.5 fm$.

With our results we can estimate the fraction of the reaction volume 
which is filled by the flux tubes of the strings. 
In the very central region with 1 fm thickness along the longitudinal
direction in the cms frame, we get $N_S\sim 420$ strings in a 
central Pb~+~Pb collision generated within the HSD approach.
The averaged string length $<l> \sim 0.5 fm$ is rather small due to 
secondary interaction of the leading quarks. Only a small amount
of strings reaches a length above 1.2 fm. 
With these results we get a string volume of
\begin{eqnarray}
V_s=\pi R_S^2 \cdot <l> \cdot N_S \approx \left\{  
\begin{array}{cc} 65 fm^3, & R_S=0.3 fm
\\ 115 fm^3, & R_S=0.4 fm
\end{array}
\right. .
\end{eqnarray}
The total volume in this central cylinder is given by
\begin{eqnarray}
V_t=\pi R_{Pb}^2 \cdot 1 fm \approx 130 fm^3,
\end{eqnarray}
where $R_{Pb}=6.6\,fm$ is the radius of the Pb nuclei.  
Comparing $V_s$ and $V_t$ illustrates that 50-75\% of the central volume
in an ultrarelativistic heavy ion collision is filled by strings in
the first few $fm / c$ assuming a string radius of 0.3 - 0.4 fm. 
The probability for a $c\bar{c}$ pair to pass the central
region thus is rather small.
This explains the strong dependence of the $J/\Psi$ results
on the string radius. 

\begin{figure}[h]
\centerline{{\psfig{figure=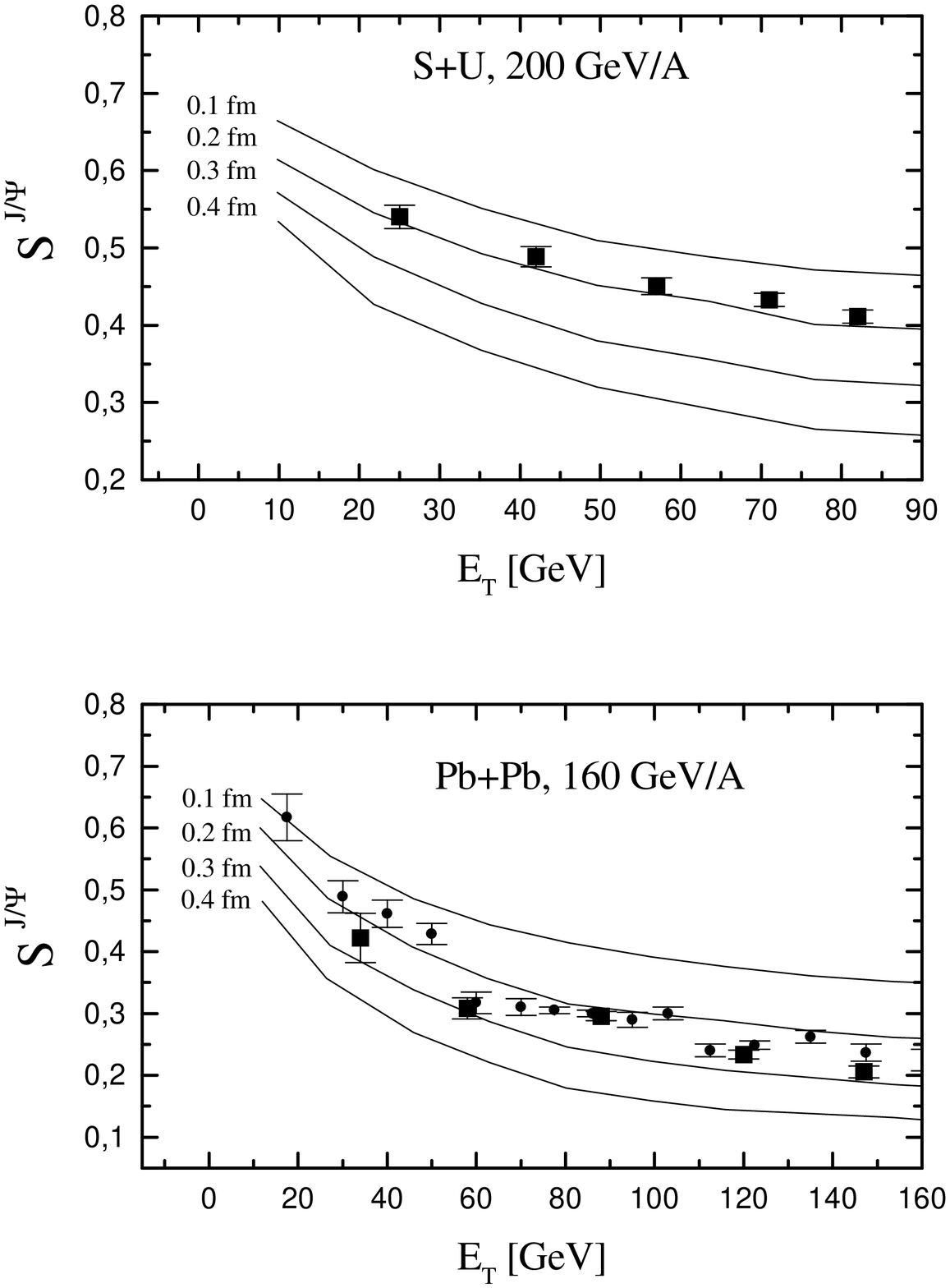,height=15cm}}}
\caption{The $J/\Psi$ survival probability $S^{J/\Psi}$ for S~+~U at 200~AGeV (upper
part) and Pb~+~Pb at 160~AGeV (lower part) as a function of the
transverse energy in comparison to the experimental data from
\protect\cite{NA50}. The circles in the lower part are the new NA50 data
from \cite{NA50a}.
Results are shown for four 
string radii from $R_S=0.1$ fm to $R_S=0.4$ fm. }
\label{hi_sup}
\end{figure}

We conclude our study by noting that $c\bar{c}$ dissociation in the 
prehadronic phase seems to be a dominant and likely charmonium  absorption 
process in relativistic heavy ion collisions. The qualitative 
behavior of the proton-nucleus and nucleus-nucleus data can be
described within this picture. 
We note that the sudden jump in the $J/\Psi$ suppression for
Pb~+~Pb suggested by the data for $E_T\approx 50\,GeV$ in Fig.
\ref{hi_sup} cannot be described in our dynamical model as in Ref.
\cite{cassing-jpsi} based on string and hadron dynamics
(cf. also \cite{kharz} for a recent overview), because the predicted
suppression within our approach is always a rather smooth function
of atomic number and centrality of the collision.
If confirmed experimentally, this might indicate a new phase with other 
degrees of freedom (partons) and reaction channels possibly
related to a deconfined QGP formation \cite{satz2}.

\newpage 

\begin{thebibliography}{99}
\bibitem{matsui}
        T. Matsui and H. Satz, Phys. Lett. B 178 (1986) 416.
\bibitem{NA38}
        NA38 Collaboration, C. Baglin et al.,
        Phys. Lett.  B 270 (1991) 105; Phys. Lett.  B 345 (1995) 617;
        S. Ramos, Nucl. Phys. A 590 (1995) 117c.
\bibitem{gonin}
        NA50 Collaboration, M. Gonin et al., Nucl. Phys.  A 610 (1996) 404c;
\bibitem{NA50}
        NA50 Collaboration, F. Fleuret  et al.,
        in '97 QCD High Energy Hadronic Interactions,
        ed. by Tran Thanh Van, Editions Fronti\'eres, (1997) 503
\bibitem{NA50a}
        NA50 Collaboration, L. Ramello, to appear in the Proc. of the 
        Quark Matter '97 Conference.
\bibitem{gerhuf}
        C. Gerschel and J. H\"ufner, hep-ph/9802245.
\bibitem{KhaQM96}
        D. Kharzeev,    Nucl. Phys. A 610 (1996) 418c.
        J. H\"ufner and BZ. Kopeliovitch, Phys. Rev. Lett. 76 (1996) 192
\bibitem{blaizot}
        J. P. Blaizot and J. Y. Ollitrault,
        Phys. Rev. Lett. 77 (1996) 1703; Nucl. Phys. A 610 (1996) 452c.
\bibitem{wong}
        C.-Y. Wong, Nucl. Phys.  A 610 (1996) 434c.
\bibitem{gerschel}
        C. Gerschel and J. H\"ufner, Z. Phys.  C 56  (1992) 71;
        C. Gerschel, Nucl. Phys. A 583 (1995) 643.
\bibitem{gavin}
        S. Gavin and R. Vogt, Nucl. Phys. B 345 (1990) 104;
        S. Gavin, H. Satz, R. L. Thews, and R. Vogt, Z. Phys. C 61 (1994) 351;
        S. Gavin, Nucl. Phys. A 566 (1994) 383c.
\bibitem{gavin1}
        S. Gavin and R. Vogt, Nucl. Phys.  A 610 (1996) 442c;
        Phys. Rev. Lett. 78 (1997) 1006.
\bibitem{Capella}
        A. Capella, A. Kaidalov, A. Kouider Akil, and C. Gerschel,
       Phys. Lett. B 393 (1997) 431.
\bibitem{cassing-jpsi}
        W. Cassing and E. L. Bratkovskaya, 
        Nucl. Phys. A 623 (1997), 570
\bibitem{armesto} 
        N. Armesto and A. Capella, hep-ph/9705275. 
\bibitem{cassing}
        W. Ehehalt and W. Cassing, Nucl. Phys. A 602  (1996) 449.
\bibitem{antico}
        H. Satz, hep-ph/9711289;
        D. Kharzeev and H. Satz, Phys. Lett. B 334 (1994) 155.
\bibitem{LUND}
        B. Anderson, G. Gustafson, Hong Pi, Z.\ Phys.\ C  57 (1993) 485.
\bibitem{loh}
        S. Loh, C. Greiner, and U. Mosel,
        Phys. Lett. B 404 (1997) 238.
\bibitem{cassing1}
        W. Cassing, W. Ehehalt, and C. M. Ko, Phys. Lett. B 363 (1995) 35,
        W. Cassing, W. Ehehalt, and I. Kralik, Phys. Lett. B 377 (1996) 5,
        E. L. Bratkovskaya, W. Cassing and U. Mosel,
        Z. Phys. C 75 (1997) 119,
        E. L. Bratkovskaya and W. Cassing, Nucl. Phys. A 619 (1997) 413
\bibitem{Wolf90}
        Gy. Wolf, G. Batko, W. Cassing, U. Mosel, K. Niita,
        and M. Sch\"afer, Nucl. Phys. A 517 (1990) 615,
        Gy. Wolf, W. Cassing, U. Mosel, Nucl. Phys. A 552 (1993) 459.
\bibitem{PDB}
        Particle Data Booklet, Phys. Rev. D 50 (1994) 1173.
\bibitem{Carlos}
        C. Louren\c co, PhD Thesis, Lisbon 1995 (unpublished).
\bibitem{Frankel} 
        S. Frankel and W. Frati, hep-ph/9710532; hep-ph/9710022
\bibitem{Tai}
        Tai An, Chao Wei Qin and Yao Xiao Xia, hep-ph/9701207 
\bibitem{kharz}
        D. Kharzeev, nucl-th/9802037 
\bibitem{KhSatz96}
        D. Kharzeev and H. Satz, Phys. Lett. B 366 (1996) 316.
\bibitem{satz2}
        D. Kharzeev, M. Nardi and H. Satz, hep-ph/9707308 
\end{thebibliography}
\end{document}